\documentclass[aps,prb,twocolumn,groupedaddress,floatfix,showpacs]{revtex4}
\usepackage{bm}
\usepackage{dcolumn}
\usepackage{graphicx}

\begin{document}

\title{Ligand field parameters and the ground state of Fe(II) phthalocyanine}
\author{M.D. Kuz'min$^1$, A. Savoyant$^2$, and R. Hayn$^2$}

\affiliation{ $^1$Institut f\"ur Materialwissenschaft, TU Darmstadt, 
                                                   64287 Darmstadt, Germany \\
   $^2$IM2NP, CNRS UMR 6242, FST, Aix-Marseille Universit\'e, 
                                          F-13397 Marseille Cedex 20, France }

\date{\today}

\begin{abstract}
A judicious analysis of previously published experimental data leads one to conclude
that the ground state of iron(II) phthalocyanine is an orbitally degenerate spin
triplet $a_{1g}^2 e_g^{\uparrow\downarrow\uparrow} b_{2g}^{\uparrow}$ ($^3E_g$).
The ligand field parameters, in relation to Racah's $C$, are approximately as
follows: $B_{20}/C=0.84$, $B_{40}/C=0.0074$. The uniqueness of this result is
demonstrated by means of a special diagram in the $B_{20}/C-B_{40}/C$ plane
(under additional conditions that $B_{44}/B_{40}=35/3$ and $B/C=0.227$). The system
is in a strong-ligand-field regime, which enables the use of single-determinant
techniques corrected for correlations within the $3d$ shell of Fe.
\end{abstract}
\maketitle

\section{Introduction}
Over several decades the interest in iron (II) phthalocyanine (FePc) has been
motivated by various applications as well as by its proximity to iron proteins. 
In more recent times FePc has become popular as a model system for core-level
spectroscopy (XAS, XMCD) studies.\cite{Miedema, Stepanow,Barto10} Certain progress
has been made in creating artificial ordered structures of FePc.\cite{Abel,Gredig}
Perhaps the most significant finding of the recent decade was the discovery of an
unquenched orbital moment of iron in FePc by means of in-field M\"ossbauer
spectroscopy\cite{Filoti} and XMCD.\cite{Barto10}

This has in turn brought about a surge of computational activity on FePc. Along
with various density-functional calculations,\cite{ReynoldsFiggis,LiaoScheiner,
MaromKronik,KHO,Brena,Nakamura,Wang,Sumimoto,Bialek} it is worth to mention 
multiplet structure calculations\cite{Thole,Miedema,Stepanow} based on a
phenomenological model. The main ingredients of the latter approach are the
Coulomb repulsion, allowed for by way of the Slater-Condon parameters, and the
crystal (ligand) field (CF) on Fe$^{2+}$. The multiplet calculations\cite{Thole,Miedema,Stepanow}
were mainly aimed at simulating x-ray absorption spectra; however, they produced
an interesting by-product. This is a map of ground states of Fe$^{2+}$ in CF
parameter space (Fig. 2 of Ref. \onlinecite{Miedema}). An early version of such
a diagram for the point group $D_{4h}$ was produced by K\"onig and Schnakig,\cite{KoenigS}
but the idea itself goes back to the classical work of Tanabe and Sugano,\cite{TS}
who dealt with the cubic symmetry. Unfortunately, in the case of $D_{4h}$ one cannot
plot but 2-dimensional sections of the 3-dimensional space of CF parameters,
the choice of these sections in Refs. \onlinecite{KoenigS} and \onlinecite{Miedema}
being rather suboptimal. Besides, the diagrams in Refs. \onlinecite{KoenigS} and 
\onlinecite{Miedema} have the disadvantage that CF parameters in energy units are
plotted on the axes, and so the diagrams depend on the Slater-Condon (or Racah)
parameters employed. As against that, the original work of Tanabe and Sugano\cite{TS}
presented the result in terms of a dimensionless ratio of the CF parameter to
Racah's $B$, which led to the celebrated series of universal diagrams. Still,
Miedema's diagrams are of interest. They have an enigmatic cornered shape, the 
domain boundaries are piecewise-linear, with repeatedly encountered, characteristic
slopes. These features of the diagrams have so far remained unexplained.

As regards agreement with experiment, the calculations leave much to be desired.
Density-functional calculations make inconclusive predictions of the ground state.
Thus, Reynolds and Figgis\cite{ReynoldsFiggis} could not decide between $^3E_g$ and
$^3B_{2g}$ because the two lie too close in energy. Marom and Kronik\cite{MaromKronik}
found either $^3B_{2g}$ ($e_g^4 a_{1g}^1 b_{2g}^1$) 
or $^3A_{2g}$ ($a_{1g}^2 b_{2g}^2 e_g^{\uparrow\uparrow} $), depending on 
computational details. \cite{remark1}
More
recently, Nakamura et al.\cite{Nakamura} found $^3A_{2g}$ in an isolated FePc
molecule, but $^3E_g$ in a columnar stack of such molecules. Establishing the symmetry
of the ground state does not settle the dispute: within the correct $^3E_g$
one should further distinguish between the configurations $b_{2g}^2 e_g^3 a_{1g}^1$, 
as conjectured by Dale et al.,\cite{Dale} and $a_{1g}^2 e_g^3 b_{2g}^1$, found in
the multiplet calculations.\cite{Miedema,Stepanow} The two ground-state 
configurations 
lead to distinct types of magnetic behavior.

This work aims at determining the CF parameters of Fe(II) phthalocyanine. 
As we will show below, the known experimental facts on that compound (obtained by
magnetic and spectroscopic measurements) in connection with our CF analysis
leave no choice: there is only one domain in the field of CF parameters yielding a 
ground state that does
not contradict established knowledge. In such a way our calculations resolve
the confusing puzzle about the ground state of FePc that existed for many years. As
a byproduct the peculiar shape of Miedema's diagrams is explained.

In the following, we consider the $3d^6$ configuration in
a CF of symmetry $D_{4h}$ and allow for Coulomb repulsion between the $3d$ 
electrons. The CF is {\em a priori} assumed neither strong nor weak as compared
with the Coulomb interaction. Hybridization of the Fe $3d$ orbitals 
to neighboring ligands is thought to be included into the relevant CF parameters.
In that sense it is better to call our theory ligand field theory instead. But
the ligand $p$ orbitals are not treated in an explicit way and we keep the term
CF theory for simplicity. The spin-orbit coupling is neglected at first (since it is much
weaker than either the CF or the Coulomb repulsion) but taken into account in a
later discussion of magnetic properties.

This paper is organized as follows. In the next section we briefly review the
experimental facts that bear on our knowledge of the ground state of FePc and
reiterate the current status of this knowledge. Further, in Section III, a diagram
of ground states of FePc is constructed from numerical calculations. Our diagram
is similar to that of Ref. \onlinecite{Miedema}, the main two differences being
that (i) dimensionless coordinates of the Tanabe-Sugano type are used and (ii)
the section of the 3-dimensional space of CF parameters is chosen on the principle
that the coordination polyhedron is a plane square. In Section IV, the same diagram
is reproduced analytically, which includes explicit expressions for all domain
boundaries. The subsequent discussion in Section V hinges upon the good agreement
of the exact (numerical) and approximate (analytical) diagrams. The piecewise-linear
shape of the domain boundaries finds a natural explanation in the linearity of the
underlying equations. A conclusion is made that FePc is in a strong CF mode and 
approximate values of the CF parameters are given (or rather, ratios of CF 
parameters to Racah's $C$). The ground-state configuration turns out to be
$a_{1g}^2 e_g^3 b_{2g}^1$ ($^3E_g$), as in Refs. \onlinecite{Miedema,Stepanow}.
Section VI recapitulates the conclusions.

\section{Experimental facts and their implications}
\subsection{Magnetic susceptibility}
As early sources of our knowledge of the ground state of FePc one usually cites
magnetic susceptibility studies of $\beta$-FePc powder\cite{Dale} and single
crystals.\cite{Barraclough} The experimental data of both papers are in 
reasonable agreement with each other. At temperatures between 100 and 300 K the
susceptibility follows the Curie-Weiss law with $\mu_{\rm eff} \approx 
3.8\,\mu_{\rm B}$ (for powder). This is between the spin-only values of $\mu_{\rm eff}$
for $S=1$ and $S=2$ ($2\sqrt{2}\,\mu_{\rm B} \approx 2.8\,\mu_{\rm B}$ and $2\sqrt{6}\,
\mu_{\rm B} \approx 4.9\,\mu_{\rm B}$, respectively). Below about 20 K the 
susceptibility of $\beta$-FePc becomes temperature-independent.

These facts found an explanation in a simple model with $S=1$ and effective
$g$-factors employed in both works.\cite{Dale,Barraclough} The spectrum of the
model consists of a singlet ground state with $M_S=0$ and an excited doublet
with $M_S=\pm 1$ situated at $\sim 70$ cm$^{-1}$. It is unclear why Dale et 
al.\cite{Dale} thought to justify this model by proposing $b_{2g}^2 e_g^3 a_{1g}^1$
($^3E_g$) as the ground configuration (and calling it an orbital singlet). Their
work contains no experimental evidence of $^3E_g$ being the ground state of FePc.
Barraclough et al.\cite{Barraclough} noticed the discrepancy between the orbitally
degenerate $^3E_g$ and Dale's assertion that the ground state should be an orbital 
singlet, and postulated $^3B_{2g}$ instead. As pointed out in Ref. \onlinecite{StillmanTh},
this was no proof, $^3A_{2g}$ could have done equally well.

The model used in Refs. \onlinecite{Dale} and \onlinecite{Barraclough} is not
without its difficulties. So, it cannot explain the presence of an excited state
(or states) at $\sim 300$ cm$^{-1}$, as pointed out in Ref. \onlinecite{Dale}.
The existence of such an excited state follows from the fact that the 
susceptibility deviates from the Curie-Weiss law above room temperature, as 
observed by Lever.\cite{Lever} (A slight downward curvature is also visible
in $\chi^{-1}$ vs $T$ data obtained more recently on $\alpha$-FePc.\cite{Filoti})
This can be viewed as an argument in favor of $^3E_g$ rather than an orbital singlet.
The six-fold degenerate $^3E_g$ would be split by the spin-orbit interaction
into 4 singlets and a doublet, the overall splitting being $\sim\zeta\sim 400$
cm$^{-1}$. The observed susceptibility behavior would find a plausible explanation
if one of the singlets was the ground state, the doublet (or a quasi-doublet)
was situated at $\sim 70$ cm$^{-1}$, and a further state (or states) at $\sim 300$
cm$^{-1}$.

Another difficulty of Dale's triplet model consists in the values of the $g$-factors,
which differ significantly from 2. Thus, Dale et al.\cite{Dale} obtain $g_{\perp}=2.86$
(and $g_{||}=1.93$). That is, nearly one Bohr magneton has to come from an orbital
moment. Such a large orbital contribution is explained more naturally by the
presence of an unquenched orbital moment (i.e., by orbital degeneracy of the ground 
state) rather than by mixing in of excited states. We note that Barraclough et 
al.,\cite{Barraclough} who assert most emphatically the equivalence of their 
approach to that of Ref. \onlinecite{Dale}, obtained an isotropic $g$-factor,
$g_{\perp}=g_{||}=2.64$. Generally speaking, Barraclough's $g$-factors should be
more trustworthy, since they were deduced from data measured on a single 
crystal.\cite{Barraclough} The difficulty, however, is that, according to Eq. (4)
of Ref. \onlinecite{Dale}, the zero-field splitting must vanish for 
$g_{\perp}=g_{||}$. At the same time, it is emphasized that this splitting,
$\sim 70$ cm$^{-1}$, is very large.\cite{Barraclough} 

In any case, it should be regarded as firmly established that the susceptibility
is a maximum in the plane of the FePc molecule.\cite{Barraclough} This conclusion
has been recently confirmed in an independent experiment.\cite{Barto10} As regards
the ground states conjectured to explain the susceptibility data, they cannot be
viewed as deduced from experiment.

\subsection{Other techniques}
An x-ray diffraction experiment of Coppens et al.\cite{Coppens} found the occupation
numbers of the Fe 3d orbitals in FePc: $b_{2g}^{1.65} e_g^{2.13} a_{1g}^{0.88}
b_{1g}^{0.75}$. On account of covalency, these numbers sum up to 5.41 rather than 6.
Restoring the normalization to 6, one has $b_{2g}^{1.83} e_g^{2.36} a_{1g}^{0.98}
b_{1g}^{0.83}$. Coppens et al. regarded their result as a direct confirmation of
Dale's conjecture, $b_{2g}^2 e_g^3 a_{1g}^1$ ($^3E_g$). Yet, the analysis in Ref.
\onlinecite{Coppens} was limited to spin-triplet states. An unprejudiced look at
the quintet states, in particular at $b_{2g}^2 e_g^{\uparrow\uparrow} 
a_{1g}^{\uparrow} b_{1g}^{\uparrow}$ ($^5B_{2g}$), suggests a higher degree of
agreement with Coppens' results. However, $^5B_{2g}$ can be ruled out because it 
would have resulted in too high a magnetic moment, $\mu_{\rm eff} = 4.9\,\mu_{\rm B}$. 

Turning now to the optical absorption experiments of Stillman and Thomson,\cite{StillmanTh}
we note that they were carried out on FePc solution in dichlorobenzene. This system
is chemically distinct from either the free FePc molecule or $\alpha$ or $\beta$
FePc. Therefore, without casting doubt upon Stillman and Thomson's assertion of
a $^3A_{2g}$ ground state, we state merely that their result is not relevant to
the system under consideration herein.

A M\"ossbauer spectroscopy study of Filoti et al.\cite{Filoti} found in $\alpha$-FePc
a very large (66 T) hyperfine field on $^{57}$Fe. Unlike the usual Fermi's contact
field, the hyperfine field in $\alpha$-FePc has a positive sign (meaning 
${\bm H}_{\rm hf} \uparrow\uparrow {\bm \mu}_{\rm Fe}$) and can only originate from 
a large unquenched orbital moment. The latter was estimated to be about 
1,\cite{Filoti} but no definite information about its orientation could be obtained.

A more recent XMCD experiment of Bartolom\'e et al.\cite{Barto10} found in FePc
an orbital moment of $0.53\,\mu_{\rm B}$ lying in the plane of the molecule. In the
same work\cite{Barto10} it was demonstrated by direct measurements that the plane
of the molecule contains the easy magnetization direction, in agreement with the
early finding of Barraclough et al.\cite{Barraclough}

To summarize the section, there is no experimental evidence of the ground state
of FePc being either $^3B_{2g}$ or $^3A_{2g}$. Nor do Coppens'
data\cite{Coppens} provide sufficient confirmation for Dale's conjecture of
$b_{2g}^2 e_g^{\uparrow\downarrow\uparrow} a_{1g}^{\uparrow}$ ($^3E_g$). All one
can say at this point is that it should be a $^3E_g$ state endowed with magnetic
anisotropy of an easy-plane kind.

\section{Numerical calculations}
\subsection{Crystal field Hamiltonian}
The CF Hamiltonian operating on a single $3d$ electron in a tetragonal ($D_{4h}$)
environment is written as follows:
\begin{equation}
{\cal H}_{\rm CF} = B_{20} O_2^0 + B_{40} O_4^0 + B_{44} O_4^4
\label{HCF}
\end{equation}
Here $O_n^m$ are Stevens' operator equivalents\cite{Stevens} in the 
$\ell$-representation ($\ell =2$): $O_2^0 = 3\ell_z^2-6$, $O_4^0 = 35\ell_z^4
-155\ell_z^2+72$, $O_4^4 = \frac{1}{2}(\ell_+^4+\ell_-^4)$; $B_{nm}$ are CF
parameters. In older literature one sometimes comes across Ballhausen's CF
parameters.\cite{Ballhausen} These are related to the $B_{nm}$'s in a simple way:
\begin{equation}
Dq = \textstyle\frac{12}{5}B_{44},~~~Ds = 3B_{20},~~~
Dt = \textstyle\frac{12}{5}B_{44} - 12B_{40}
\label{Ballh}
\end{equation}
It is well known that the five real $d$ orbitals belong to distinct irreducible 
representations of the point group $D_{4h}$. Therefore, in the basis of those 
orbitals the CF Hamiltonian (\ref{HCF}) takes a diagonal form, the eigenvalues
being\cite{Ballhausen}
\begin{equation}
\begin{array}{lcccl} 
E(d_{xy})     & = &   E(b_{2g})  & = &  6 B_{20} + 12 B_{40} - 12 B_{44}     \\
E(d_{xz,yz})  & = &   E(e_g)     & = & -3 B_{20} - 48 B_{40}                 \\
E(d_{z^2})    & = &   E(a_{1g})  & = & -6 B_{20} + 72 B_{40}                 \\
E(d_{x^2-y^2})& = &   E(b_{1g})  & = &  6 B_{20} + 12 B_{40} + 12 B_{44}
\end{array}
\label{d_energies}
\end{equation}
Note that Ballhausen's original equations (5-14) and (5-15) need to be augmented 
with the cubic terms, $+6Dq$ and $-4Dq$, respectively, before being converted to 
the Stevens notation by means of Eqs. (\ref{Ballh}).

So far no restrictions have been imposed on the CF, except that it should be
compatible with the point group $D_{4h}$. Yet, much more is known about the
structure of the FePc molecule than just the symmetry of the Fe site. Thus,
the nearest environment of the iron atom consists of four nitrogen atoms making
a plane square, the Fe-N bonds being aligned with either the $x$ or the $y$ axis.
This fact enables us to reduce the number of independent CF parameters by one.
A rather general CF model known as the superposition model (see Ref. 
\onlinecite{NewmanNg} for a comprehensive review), relates pairs of CF parameters
$B_{nm}$ with equal $n$ on the basis of shape of the coordination polyhedron.
Omitting the rather straightforward calculations, we state the result: for a plane
square the superposition model demands that
\begin{equation}
B_{44} = \textstyle\frac{35}{3} B_{40}
\label{superpos}
\end{equation}

\subsection{Hamiltonian matrix}
Our calculations dealt with a Hamiltonian consisting of the CF (\ref{HCF}) and 
the Coulomb repulsion and operating on the $3d^6$ configuration. The basis states
were taken in the form of simple products of one-electron $d$ orbitals,
\begin{equation}
\prod_{i=1}^6 | m_i \sigma_i \rangle
\label{prod}
\end{equation}
with $m_i =0,\pm 1,\pm 2$, and $\sigma_i = \pm 1/2$. There are ${10 \choose 6} 
= 210$ such states in total.

Nonzero matrix elements of ${\cal H}_{\rm CF}$ (\ref{HCF}) are of two kinds. First
of all, there are diagonal matrix elements, given by
\begin{equation}
B_{20} \sum_{i=1}^6 \left( 3m_i^2 - 6 \right) +
B_{40} \sum_{i=1}^6 \left( 35m_i^4 - 155m_i^2 + 72 \right)
\label{diagCF}
\end{equation}
Secondly, there are nonzero matrix elements between the states (\ref{prod}) that
differ in one pair of quantum numbers $m_i$, $m_i$ being $-2$ in one of the states
and $+2$ in the other one. All such matrix elements equal $12B_{44}$.

The matrix elements of the Coulomb repulsion have been treated extensively in the
literature. Here we follow Griffith's fundamental treatise.\cite{Griffith} Again,
there are two distinct kinds of nonzero matrix elements. The diagonal ones are
given by
\begin{equation}
\sum_{k=0,2,4} F^k \sum_{i>j} \left[ c_{m_i m_i}^k c_{m_j m_j}^k
-\delta_{\sigma_i \sigma_j} \left( c_{m_i m_j}^k \right)^2 \right]
\label{diagCoul}
\end{equation}
where $F^k$ are the Slater-Condon parameters ($k=0,2,4$) and
\begin{equation}
c_{mm'}^k = \sqrt{\frac{4\pi}{2k+1}} \int Y_{2m}^* Y_{2m'} Y_{k,m-m'} d\Omega
\label{Gaunt}
\end{equation}
The integral in Eq. (\ref{Gaunt}) is known as the Gaunt coefficient. Numerical
values of $c_{mm'}^k$ were taken from Table 4.4 of Griffith's book.\cite{Griffith} 
The inner sum in Eq. (\ref{diagCoul}) is taken over all 15 pairs of filled $d$
orbitals. The first term in brackets describes the so-called Coulomb contribution,
while the second one, relevant to pairs of orbitals with parallel spins only,
is the exchange contribution.

The Coulomb repulsion also has off-diagonal matrix elements. These are nonzero
only between the states with equal $M_L$ and $M_S$ that differ in two occupied
$d$ orbitals, say, $|m_{i1}\sigma_{i1}\rangle$ and $|m_{j1}\sigma_{j1}\rangle$
in State \#1, as against $|m_{i2}\sigma_{i2}\rangle$ and $|m_{j2}\sigma_{j2}
\rangle$ in State \#2. It must hold that $m_{i1}+m_{j1} = m_{i2}+m_{j2}$ and
$\sigma_{i1}+\sigma_{j1} = \sigma_{i2}+\sigma_{j2}$. The matrix element between
the states \#1 and \#2 is expressed as follows
$$  \sum_{k=0,2,4} F^k \left[ \delta_{\sigma_{i1}\sigma_{i2}} 
\delta_{\sigma_{j1}\sigma_{j2}}  c_{m_{i1}m_{i2}}^k  c_{m_{j2}m_{j1}}^k \right. $$
\begin{equation}
\left. -\delta_{\sigma_{j1}\sigma_{i2}} \delta_{\sigma_{i1}\sigma_{j2}}
               c_{m_{j1}m_{i2}}^k   c_{m_{j2}m_{i1}}^k  \right]
\label{offdiagCoul}
\end{equation}
Thus, the matrix elements of the Hamiltonian are linear combinations of the CF
parameters, $B_{20}$, $B_{40}$, and $B_{44}$, as well as the Slater-Condon
parameters, $F^0$, $F^2$, and $F^4$. The latter are conveniently replaced by the
Racah parameters,
\begin{equation}
F^0 = A +\textstyle\frac{7}{5}C, ~~~F^2 = 49B+7C,~~~F^4 = \textstyle\frac{63}{5} C
\label{defRacah}
\end{equation}
The parameter $A$ is hereafter set to zero, because its only effect is to shift
the energies of all the states of $d^n$ by the same amount, $An(n-1)/2$.

\subsection{Degeneracy diagram}
The calculation consisted in setting and numerically diagonalizing the Hamiltonian
matrix for given sets of parameters $B_{20}$, $B_{40}$, $B_{44}$, $B$, $C$, and
subsequently determining the degeneracy of the ground state. Five characteristic
values of degeneracy were encountered:
$$
\begin{array}{rcl}
1: &~~~~^1A~~~~& S=0,~{\rm no~orbital~degeneracy}     \\
3: &    ^3A    & S=1,~{\rm no~orbital~degeneracy}     \\
5: &    ^5A    & S=2,~{\rm no~orbital~degeneracy}     \\
6: &    ^3E    & S=1,~{\rm double~orbital~degeneracy} \\
10: &   ^5E    & S=2,~{\rm double~orbital~degeneracy}
\end{array}
$$
At this stage the ground states are labeled tentatively. So $A$ can be anything
of the following: $A_{1g}$, $A_{2g}$, $B_{1g}$, or $B_{2g}$, which we are unable 
to distinguish. On the other hand, $^3E$=$^3E_g$ and $^5E$=$^5E_g$ as will be explained
in Section IV. 

\begin{figure}
\centerline{\includegraphics[width=0.47\textwidth]{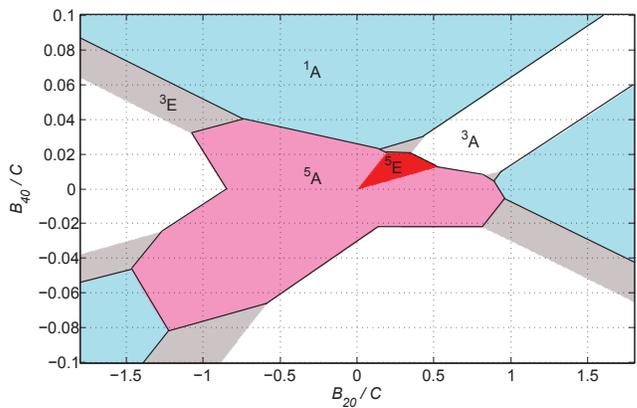}}
\caption{\label{numerical} Partition of CF parameter space among differently
degenerate ground states, as found from numerical calculations in the absence
of spin-orbit interaction. The possible ground states are denoted according to their total spin 
and the absence ($^{2S+1}A$) or presence ($^{2S+1}E$) of orbital degeneracy.
They will be further specified in Figure \ref{analytic}.}
\end{figure}

The construction of the diagram (Figure \ref{numerical}) was organized as follows. 
All energies were expressed in the units of the Racah parameter $C$. The ratios 
$B_{20}/C$ and $B_{40}/C$ were treated as independent variables defined on a dense 
mesh. In the spirit of Tanabe and Sugano,\cite{TS} the ratio $B/C$ was fixed to
a value appropriate for Fe$^{2+}$, $B/C = 0.227$, as in Table 7.3 of Ref. 
\onlinecite{AB}. The CF parameter $B_{44}$ was not regarded as an independent one. 
Rather, it was found from Eq. (\ref{superpos}), as prescribed by the superposition 
model.\cite{NewmanNg} As a result, the $B_{20}/C-B_{40}/C$ plane was partitioned 
into domains of five different kinds, according to the degeneracy found at each 
point. The diagram (Figure \ref{numerical}) has a cornered shape reminiscent of
the diagrams in Refs. \onlinecite{Miedema} and \onlinecite{KoenigS}. The domain
boundaries appear straight lines with characteristic slopes. Several sets of
parallel lines are encountered. The central part of the diagram is an area of
weak CF; in compliance with Hund's first rule, the ground state here has $S=2$.
The periphery of Figure \ref{numerical} is a region of strong CF; here $S=0$ or 1.

The numerical calculations have the advantage of producing an immediate graphical
result. However, it is not easy to analyze the character of a ground state 
expressed in a 210-dimensional basis. Of special interest to us is $^3E_g$, which
appears in six non-adjacent domains in Figure \ref{numerical}. So we would like
to know if those $^3E_g$ are similar or distinct. Furthermore, we would like to
find out the origin of the cornered shape of the diagram, why the boundaries are
straight and the slopes repeated. Finally, we pose a question, how the diagram
would change if $B/C$ and/or $B_{44}/B_{40}$ were different to the ones used so
far. Answers to the above questions should be sought by means of analytical
calculations.

\section{Analytical treatment}
\subsection{Weak crystal field: quintet states}
In the weak-field approximation the CF is treated as a perturbation with respect 
to the intra-atomic Coulomb interaction, whose eigenstates are spectral terms with
certain $L$ and $S$. Since the CF acts on spatial but not on spin variables, terms
with different $S$ do not mix together (as long as the spin-orbit coupling is 
neglected). In the $d^6$ configuration there is a single quintet term, $^5D$,
whose Coulomb energy is\cite{Griffith}
\begin{equation}
E_{\rm Coulomb} = -35B + 7C
\label{Coulomb5D}
\end{equation}

The remaining task consists in diagonalizing the CF Hamiltonian (\ref{HCF}) on the
wave functions of $^5D$, since there are no other terms with $S=2$. To this end,
it is convenient to interpret Eq. (\ref{HCF}) in a slightly different way than it
was done in Section III. Namely, $O_n^m$ are now regarded as Stevens' operators
in the $L$ representation ($L=2$): $O_2^0 = 3L_z^2 - 6$ etc. Since $^5D$ contains
a single $d$ electron above a closed semi-shell, it is only this one electron that
is exposed to the CF. Therefore, $L=\ell$ and the coefficients $B_{nm}$ in Eq.
(\ref{HCF}) are the same in both representations. So we can simply take over the
one-electron CF energies (\ref{d_energies}). In doing so, we capitalize the irrep
labels, to indicate that they now refer to many-electron states, and append the
multiplicity 5. We also prefix $E_{\rm Coulomb}$ (\ref{Coulomb5D}). The resulting 
energies of the quintet states are as follows:
$$
\begin{array}{lll}
E(^5B_{2g})  & = -35B + 7C + 6 B_{20} + 12 B_{40} - 12 B_{44} & ~({\rm Q1})  \\
E(^5E_g)     & = -35B + 7C - 3 B_{20} - 48 B_{40}             & ~({\rm Q2})  \\
E(^5A_{1g})  & = -35B + 7C - 6 B_{20} + 72 B_{40}             & ~({\rm Q3})  \\
E(^5B_{1g})  & = -35B + 7C + 6 B_{20} + 12 B_{40} + 12 B_{44} & ~({\rm Q4})
\end{array}
$$

\subsection{Strong crystal field: singlet states}
\begin{figure}
\centerline{\includegraphics[width=0.47\textwidth]{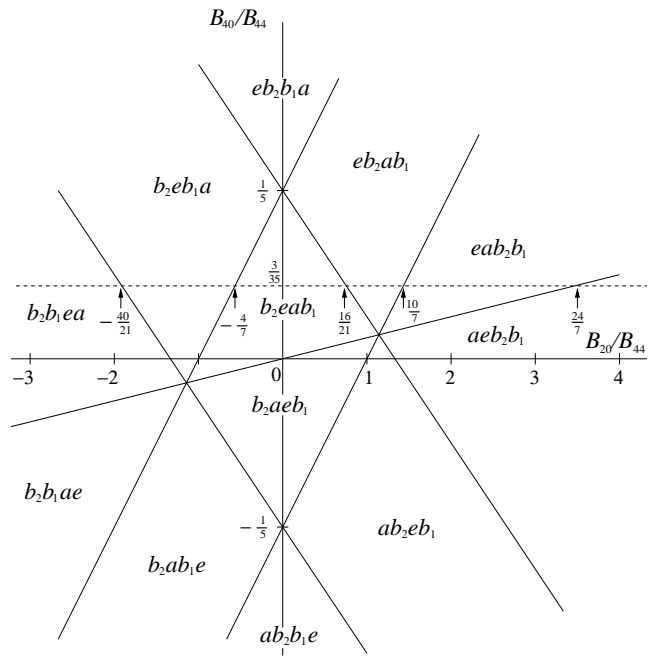}}
\caption{\label{fig2} Partition of the $B_{20}/B_{44}-B_{40}/B_{44}$ plane among all
possible permutations of the four one-electron CF levels. }
\end{figure}

In the strong-CF approximation the zeroth-order states are constructed from
one-electron eigenstates of the CF Hamiltonian (\ref{HCF}), then their energies
are corrected for the Coulomb repulsion. A first question that arises is: which
six one-electron $d$ states are filled in a CF of symmetry $D_{4h}$? To give
a possibly general answer, it is convenient to express all relevant energies in
the units of $B_{44}$. Thus, the one-electron CF energies (\ref{d_energies}) are
divided by $B_{44}$. Then, equating pairs of the so modified expressions, one 
obtains 5 equations linear in $B_{20}/B_{44}$ and $B_{40}/B_{44}$. The corresponding
lines in the parameter plane $B_{20}/B_{44} - B_{40}/B_{44}$ (Figure \ref{fig2})
are loci of points where the sequence of CF levels changes. For example, the levels
$a_{1g}$ and $b_{1g}$ cross over on a line decribed by
\begin{equation}
-B_{20}/B_{44} + 5 B_{40}/B_{44} = 1
\label{example}
\end{equation}
as readily obtained by equating the last two Eqs. (\ref{d_energies}). Equation 
(\ref{example}) describes the upper one of the two parallel lines in Figure \ref{fig2};
the lower line arises from the condition $E(a_{1g}) = E(b_{2g})$. Likewise, the
equation $E(e_g) = E(b_{1g,2g})$ generates a pair of parallel lines with a negative
slope. A single line passing through the origin is produced by the relation $E(a_{1g})
=E(e_g)$. Finally, the equation $E(b_{1g}) = E(b_{2g})$ leads to no line; this is why
there are 5 solid lines in Figure \ref{fig2}, rather than 6 as expected
combinatorially. The CF levels $b_{1g}$ and $b_{2g}$ do not swap at any inner point
of Figure \ref{fig2}, but do so at infinity, where $B_{44}$ changes sign. Therefore,
$b_2$ (a short for $b_{2g}$) stands always to the left of $b_1$ ($b_{1g}$) in the
level sequences indicated within each one of the 12 domains. The sequence labels,
read from left to right, name the CF levels in order of ascending energy if $B_{44}>0$,
and in order of descending energy if $B_{44}<0$.

Up to this point no restrictions have been placed on the CF, apart from those imposed
by the $D_{4h}$ symmetry. Now we do restrict the CF by demanding that it must
additionally comply with the superposition model, Eq. (\ref{superpos}). This implies
that the system is now bound to the horizontal dashed line in Figure \ref{fig2}.
The dashed line cuts through six domains. The corresponding intervals on the abscissa
axis are numbered 1 to 6 in order of ascending $B_{20}/B_{44}$, with $B_{44}>0$.
Thus, the interval \#1 stands for $B_{20}/B_{44} <-\frac{40}{21}$, \#2 means that
$-\frac{40}{21} < B_{20}/B_{44} < -\frac{4}{7}$ etc. The same intervals, but with
$B_{44}<0$, will be referred to by overscore numbers $\overline{1}$ to $\overline{6}$.

Now, examining the CF level sequences in the above 12 intervals, one encounters only
3 situations where there is a CF gap between the highest occupied and the lowest 
unoccupied orbitals. The corresponding ground-state electronic configurations are as
follows:
\begin{equation}
\begin{array}{ll}
b_{2g}^2 e_g^4               &   {\rm ~intervals}~\#2,\,\#3,\,\#4                \\
a_{1g}^2 e_g^4               &   {\rm ~intervals}~\#5,\,\#6,\,\#\overline{1}     \\
a_{1g}^2 b_{1g}^2 b_{2g}^2   &   {\rm ~intervals}~\#\overline{4},\,\#\overline{5}
\end{array}
\label{intervals}
\end{equation}

Within the above intervals of $B_{20}/B_{44}$ the ground state is a singlet, provided
that the CF is sufficiently strong. In all other cases the Fermi level is caught at
the partially occupied quadruply degenerate $e_g$ level and there is a possibility
of both a singlet and a (spin) triplet ground state with the same energy.
A subsequent allowance for intra-atomic (Hund's) exchange makes the triplet states
energetically more favorable than the singlet ones. Therefore, singlets are no viable
candidates for ground state in all intervals where triplets with the same CF energy
are possible. In such intervals only the triplets will be considered (in the next
subsection).

Conversely, in the three cases where there are viable singlet states (\ref{intervals}),
competing triplet states will be taken into consideration as well, constructed from
excited CF configurations. Such triplets still have a chance of becoming ground state
on account of Hund's exchange in situations where the CF is not strong enough, near
interval boundaries, etc.

Let us now turn to our direct task --- computing the energies of the singlet states
(\ref{intervals}). The CF energies are computed most readily, by summing up the 
energies of the six occupied one-electron states as given by Eqs. (\ref{d_energies}).
First-order correlation corrections are then computed following Slater's 
prescription:\cite{Slater} for each pair of occupied $d$ states a so-called Coulomb
integral $J(d_1,d_2)$ is added; a further {\em exchange} contribution $K(d_1,d_2)$
is deducted for pairs with equal spins. $J$'s and $K$'s between the real $d$ orbitals
were expressed in terms of the Racah parameters by Griffith, see Table A26 of his
book.\cite{Griffith} The resulting singlet energies are as follows:
$$
\begin{array}{lcrl}
E(b_{2g}^2e_g^4)            & = & -168 B_{40} - 24 B_{44} -30 B + 15 C  & ~~({\rm S1})  \\
E(a_{1g}^2e_g^4)            & = & -24 B_{20} - 48 B_{40} + 10 B + 15 C  & ~~({\rm S2})  \\
E(a_{1g}^2b_{1g}^2b_{2g}^2) & = & 12 B_{20} + 192 B_{40} - 20 B + 15 C  & ~~({\rm S3})
\end{array}
$$
We note that for low-lying single-product states, such as those considered in this work,
the factor of $C$ depends solely on $S$ and is given by
\begin{equation}
({\rm factor~of~}C) = (S_{\max}+1)^2 - (S+1)^2
\label{factorofC}
\end{equation}
where $S_{\max}=n/2$ is the hypothetical maximum spin of $n$ electrons
in the absence of the Pauli principle. For fewer than six $d$ electrons, states with
$S=S_{\max}$ are allowed and their energies have no contribution in $C$. For $d^6$,
$S_{\max}=3$ and the factors of $C$ equal 15, 12, and 7 for $S=0$, 1, and 2, 
respectively. Equation (\ref{factorofC}) is a consequence of the great simplicity
acquired by the Coulomb and exchange integrals when $A$ and $B$ are set to zero:
$$
J(d_i,d_j) = K(d_i,d_j) = (1+2\delta_{ij}) C,
$$
cf. Table A26 of Ref. \onlinecite{Griffith}. No simple relations are known for the
factors of $B$, which have to be calculated in each case separately.

\subsection{Strong crystal field: triplet states}
Construction and finding the energies of the (spin) triplet states are carried out in
a similar fashion. One peculiarity is the large number of triplets (9 in total),
which have to be constructed for all 12 intervals of $B_{20}/B_{44}$. Where no triplet 
state is permitted by the ground CF configuration, the first excited configuration
will be considered instead.

We proceed from the interval \#1, $B_{20}/B_{44} < -\frac{40}{21}$, $B_{44}>0$. Here
(as well as in the interval \#$\overline{6}$, $B_{20}/B_{44} > \frac{24}{7}$, 
$B_{44}<0$) the ground CF configuration is $b_{1g}^2 b_{2g}^2 e_g^2$, which allows
one triplet state, $d_{x^2-y^2}^2 d_{xy}^2 d_{xz}^{\uparrow} d_{yz}^{\uparrow}$,
as well as three singlet ones. According to the first Hund's rule, it will be the
triplet that will become ground state upon allowance for the Coulomb interaction.
(It was for this reason that the singlets were left out in the previous subsection.)
The symmetry of the triplet state is $^3A_{2g}$, as determined
by the antisymmetrized product of $d_{xz}$ and $d_{yz}$.
The energy is computed following the same
prescription as in the previous subsection and equals
$$
E(b_{1g}^2 b_{2g}^2 e_g^{\uparrow\uparrow}) = 18 B_{20} - 48 B_{40} - 9B + 12C 
                                                                 ~~~~~~({\rm T1})
$$

Let us move to the interval \#2, $-\frac{40}{21} < B_{20}/B_{44} < -\frac{4}{7}$,
$B_{44}>0$. The ground CF configuration, $b_{2g}^2e_g^4$, consists of fully occupied
orbitals and is necessarily a singlet. To construct a spin triplet state, one spin-down
electron is promoted, with a simultaneous reversal of spin, from the $e_g$ orbital
to the first unoccupied CF level $b_{1g}$. The result is either $d_{xy}^2 d_{xz}^2
d_{yz}^{\uparrow} d_{x^2-y^2}^{\uparrow}$ or $d_{xy}^2 d_{yz}^2 d_{xz}^{\uparrow} 
d_{x^2-y^2}^{\uparrow}$. This is a doubly orbitally degenerate state $^3E_g$. Its
energy is
$$
E(b_{2g}^2 e_g^{\uparrow\downarrow\uparrow} b_{1g}^{\uparrow})
     = 9 B_{20} - 108 B_{40} - 12 B_{44} - 24 B + 12 C         ~~~ ({\rm T2})
$$
Proceeding as before, we find that the most favorable spin triplet state in the 
interval \#3 is another $^3E_g$, whose energy is given by
$$
E(b_{2g}^2 e_g^{\uparrow\downarrow\uparrow} a_{1g}^{\uparrow})
     = -3 B_{20} - 48 B_{40} - 24 B_{44} - 28 B + 12 C         ~~~ ({\rm T3})
$$

The remaining six spin-triplet states include: a $^3B_{2g}$ in the intervals \#4 and \#5,
with
$$
E(e_g^4 b_{2g}^{\uparrow} a_{1g}^{\uparrow})
     = -12 B_{20} - 108 B_{40} - 12 B_{44} - 22 B + 12 C       ~~~ ({\rm T4})
$$
a $^3E_g$ in the interval \#6, with
$$
E(a_{1g}^2 e_g^{\uparrow\downarrow\uparrow} b_{2g}^{\uparrow})
     = -15 B_{20} + 12 B_{40} - 12 B_{44} - 14 B + 12 C        ~~ ({\rm T5})
$$
a $^3E_g$ in the interval \#$\overline{1}$, with
$$
E(a_{1g}^2 e_g^{\uparrow\downarrow\uparrow} b_{1g}^{\uparrow})
     = -15 B_{20} + 12 B_{40} + 12 B_{44} - 14 B + 12 C        ~~ ({\rm T6})
$$
a $^3A_{2g}$ in the intervals \#$\overline{2}$ and \#$\overline{3}$, with
$$
E(a_{1g}^2 b_{1g}^2 e_g^{\uparrow\uparrow})
     = -6 B_{20} + 72 B_{40} + 24 B_{44} - 29 B + 12 C         ~~~ ({\rm T7})
$$
a $^3E_g$ in the interval \#$\overline{4}$, with
$$
E(a_{1g}^2 b_{1g}^2 b_{2g}^{\uparrow} e_g^{\uparrow})
     = 3 B_{20} + 132 B_{40} + 12 B_{44} - 29 B + 12 C         ~~ ({\rm T8})
$$
and a $^3E_g$ in the interval \#$\overline{5}$, with
$$
E(b_{1g}^2 b_{2g}^2 a_{1g}^{\uparrow} e_g^{\uparrow})
     = 15 B_{20} + 72 B_{40} - 13 B + 12 C                     ~~~~ ({\rm T9})
$$
Note that the factor of $C$ in Eqs. (T1--T9) is invariably 12, as follows from Eq.
(\ref{factorofC}) with $S=1$ and $S_{\max}=3$.

\subsection{The $B_{20}/C - B_{40}/C$ diagram}
The search for the ground state consists in a systematic comparison of energies of
pairs of candidate states, as given by Eqs. (Q1-Q4, S1-S2, T1-T9). For example,
equating (Q1) to (T1) results in
\begin{equation}
-12 B_{20} + 60 B_{40} - 12 B_{44} = 26 B + 5 C
\label{Q1T1}
\end{equation}
Eliminating $B_{44}$ by means of Eq. (\ref{superpos}) and dividing the result by $C$,
one arrives at an equation of a straight line in the plane of the parameters $B_{20}/C$
and $B_{40}/C$:
\begin{equation}
\frac{B_{40}}{C} = -0.15 \frac{B_{20}}{C} - 0.0625 - 0.325 \frac{B}{C}
\label{Q1T1_bis}
\end{equation}
Left of this line there should be a domain where the ground state is the triplet $T_1$
($^3A_{2g}$ or $b_{1g}^2 b_{2g}^2 e_g^{\uparrow\uparrow}$), right of the line, towards 
the origin, lies the domain where the ground state is the quintet $Q_1$ ($^5B_{2g}$). 
In the spirit of Tanabe-Sugano, the ratio $B/C$ is fixed, $B/C=0.227$, as in Table 7.3 
of Ref. \onlinecite{AB}.
 
\begin{figure}
\centerline{\includegraphics[width=0.49\textwidth]{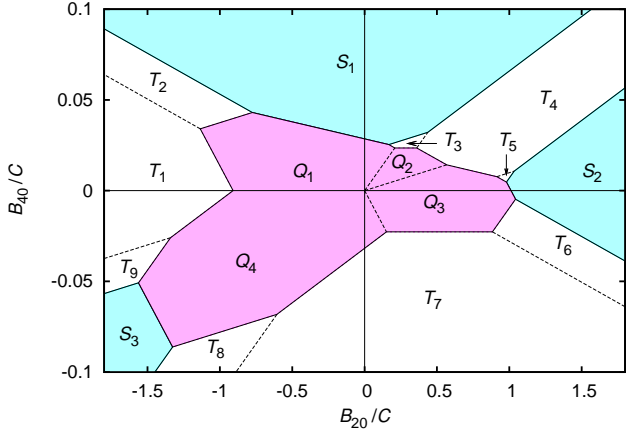}}
\caption{\label{analytic} Partition of CF parameter space among possible ground states
of FePc. The labels are mnemonic: so $S_1$ is a singlet whose energy is given by Eq. (S1).
The borderlines are as described by Eq. (\ref{A_1}) with the coefficients of Table I.}
\end{figure}

Proceeding as above, one obtains equations for all 33 borderlines appearing in the 
diagram, Figure \ref{analytic}. (It suffices to consider pairs of states belonging to
the same, or perhaps, to adjacent intervals of $B_{20}/B_{44}$.) By analogy with Eq. 
(\ref{Q1T1_bis}), these expressions are presented as
\begin{equation}
\frac{B_{40}}{C} = a \frac{B_{20}}{C} + b + b' \frac{B}{C}
\label{A_1}
\end{equation}
The numerical factors $a$, $b$, and $b'$ are listed in Table I. A line is referred to by 
naming the two domains it separates. Remarkably, one finds in the second column of Table I
repeatedly six characteristic slopes,
\begin{equation}
0,-\textstyle\frac{9}{200},-\frac{3}{20},\frac{9}{80},\frac{3}{50},{\rm ~and~}\frac{1}{40}.
\label{slopes}
\end{equation}
These are obtained by means of Eq. (\ref{superpos}) from the interval boundaries in
Figure \ref{fig2}: so $B_{20}/B_{44} = 24/7$ leads to $B_{20}/B_{40} = 1/40$ etc.
One exception that is not on the list (\ref{slopes}) but is encountered twice in the second 
column of Table I is $-3/160$.

\begin{table}
\caption{Values of coefficients in Eq. (\ref{A_1}).}
\begin{center}
\renewcommand{\arraystretch}{1.2}
\begin{tabular}{rrrr}
\hline \hline
label         &  $a~~~$        &  $b~~~$       &  $b'~~~~~~$      \\ \hline
$T_1Q_1$      &  $-3/20$       &  $-1/16$      &  $-13/40~~~$     \\
$T_2Q_1$      &  $1/40$        &  $1/24$       &  $11/120~~~$     \\
$T_1T_2$      &  $-9/200$      &  $0$          &  $-3/40~~~$      \\
$~~~~S_1T_2$  &  $~~~~-9/200$  &  $~~~~~3/200$ &  $~~~~-3/100~~~$ \\
$S_1Q_1$      &  $-3/160$      &  $1/40$       &  $1/64~~~$       \\
$S_1T_3$      &  $1/40$        &  $1/40$       &  $-1/60~~~$      \\
$S_1T_4$      &  $3/50$        &  $3/200$      &  $-1/25~~~$      \\
$T_3Q_1$      &  $-9/200$      &  $1/40$       &  $7/200~~~$      \\
$T_3Q_2$      &  $0$           &  $1/56$       &  $1/40~~~$       \\
$T_4Q_2$      &  $-9/200$      &  $1/40$       &  $13/200~~~$     \\
$T_4Q_3$      &  $-3/160$      &  $1/64$       &  $13/320~~~$     \\
$T_5Q_3$      &  $-9/200$      &  $1/40$       &  $21/200~~~$     \\
$S_2Q_3$      &  $-3/20$       &  $1/15$       &  $3/8~~~$        \\
$S_2T_5$      &  $9/80$        &  $-3/80$      &  $-3/10~~~$      \\
$S_2T_4$      &  $3/50$        &  $-3/200$     &  $-4/25~~~$      \\
$Q_1Q_2$      &  $9/80$        &  $0$          &  $0~~~$          \\
$Q_2Q_3$      &  $1/40$        &  $0$          &  $0~~~$          \\
$T_3T_4$      &  $9/80$        &  $0$          &  $-3/40~~~$      \\
$T_4T_5$      &  $1/40$        &  $0$          &  $-1/15~~~$      \\
$T_6Q_3$      &  $9/80$        &  $-1/16$      &  $-21/80~~~$     \\
$T_7Q_3$      &  $0$           &  $-1/56$      &  $-3/140~~~$     \\
$T_7Q_4$      &  $3/50$        &  $-1/40$      &  $-3/100~~~$     \\
$S_2T_6$      &  $-9/200$      &  $3/200$      &  $3/25~~~$       \\
$T_6T_7$      &  $-9/200$      &  $0$          &  $3/40~~~$       \\
$Q_3Q_4$      &  $-3/20$       &  $0$          &  $0~~~$          \\
$T_8Q_4$      &  $1/40$        &  $-1/24$      &  $-1/20~~~$      \\
$T_7T_8$      &  $9/80$        &  $0$          &  $0~~~$          \\
$S_3Q_4$      &  $-3/20$       &  $-1/5$       &  $-3/8~~~$       \\
$S_3T_8$      &  $9/80$        &  $3/80$       &  $9/80~~~$       \\
$T_9Q_4$      &  $9/80$        &  $1/16$       &  $11/40~~~$      \\
$S_3T_9$      &  $1/40$        &  $-1/40$      &  $7/120~~~$      \\
$T_1Q_4$      &  $3/50$        &  $1/40$       &  $13/100~~~$     \\
$T_1T_9$      &  $1/40$        &  $0$          &  $1/30~~~$       \\ \hline \hline
\end{tabular}
\label{tab:fit&}
\end{center}
\end{table}

\section{Discussion}
In the preceeding section we constructed a diagram of ground states of FePc in the
absence of spin-orbit coupling, Figure \ref{analytic}. A total of 16 distinct ground 
states are present in the diagram: 3 singlets ($S_1-S_3$), 9 spin triplets ($T_1-T_9$), 
and 4 spin quintets ($Q_1-Q_4$). The respective energies are given by Eqs. (S1-S3, T1-T9,
Q1-Q4). Explicit expressions were derived for the domain boundaries, Eq. (\ref{A_1}) and
Table I. The boundaries are segments of straight lines, which is a consequence of the
linearity of Eqs. (S1-S3, T1-T9, Q1-Q4). This gives the diagram
its peculiar cornered shape, with characteristic, repeated slopes. As clear from the
structure of Eq. (\ref{A_1}), the slopes do not depend on the ratio $B/C$.
Taking a slightly different $B/C$ would shift the domain boundaries somewhat, but 
will not affect their slopes. As against that, the slopes will change if the ratio
$B_{44}/B_{40}$ deviates from the value prescribed by the superposition model, Eq.
(\ref{superpos}). Moreover, such a deviation of $B_{44}/B_{40}$ from 35/3 may lead
to a loss of parallelity of certain boundary lines. For example, from a simple
analysis of the one-electron CF energies (\ref{d_energies}) one finds
$$
\begin{array}{l}
({\rm slope~of~}T_1Q_1) = ({\rm slope~of~}Q_3Q_4) = (5-B_{44}/B_{40})^{-1}    \\
({\rm slope~of~}S_3Q_4) = (2B_{44}/B_{40} -30)^{-1}                           \\
({\rm slope~of~}S_2Q_3) = -3/20
\end{array}
$$
Apparently the above lines are only parallel if the condition (\ref{superpos}) is
fulfilled. In reality the superposition model is an approximation and small deviations
from Eq. (\ref{superpos}) are to be expected. In the above example, the borderlines
$T_1Q_1$ and $Q_3Q_4$ will remain parallel exactly, while the others only 
approximately. A more extensive analysis of this matter is beyond the scope of the
present work.

On the whole, the diagram constructed analytically (Figure \ref{analytic}) is remarkably 
similar to that calculated numerically (Figure \ref{numerical}). We take it as a sign of 
validity of the strong-CF approximation used to compute the energies of the singlet and 
triplet states, Eqs. (S1-S3, T1-T9). (N.B. The quintet energies (Q1-Q4) are essentially
exact, without relying on the weakness of the CF.) This demonstrates the applicability
of techniques based on single-determinant wave functions, even though it is important
to allow for correlations (nonzero $B$ and $C$).

Our next task is to locate the standpoint of FePc in the diagrams of Figures 
\ref{numerical} and \ref{analytic}. In the subsequent discussion the domain boundaries 
are assumed to be positioned as in the more accurate Figure \ref{numerical}, whereas 
the ground states associated with the domains are as constructed analytically and 
indicated in Figure \ref{analytic}. The search can be limited to an acute angle adjacent 
to the abscissa axis, within
the first quadrant of Figure \ref{analytic}:
\begin{equation}
0 < B_{40} < 0.45 B_{20}
\label{condition}
\end{equation}
Indeed, the $d_{x^2-y^2}$ orbital of Fe
overlaps most strongly with the ligand orbitals and therefore has a much higher energy
than the other $3d$ orbitals, in particular, $d_{xy}$. By Eqs. (\ref{d_energies}),
$E(d_{x^2-y^2}) - E(d_{xy}) = 24B_{44} > 0$, whence by Eq. (\ref{superpos}), 
$B_{40}>0$. To prove the right-hand part of the double inequality (\ref{condition}),
one should rewrite the CF Hamiltonian (\ref{HCF}), taken in conjunction with Eq.
(\ref{superpos}), as a classical anisotropy energy,
$$   E_a = B_{20} (3\cos^2 \theta -1)  ~~~~~~~~~~~~~~~~~~~~~~~~~~ $$
$$   + B_{40} (35\cos^4 \theta -30\cos^2 \theta + 3 
     + \textstyle\frac{35}{3} \sin^4 \theta \cos 4\phi )          $$
and demand that $\theta =\pi /2$, $\phi = \pi /4$ be a local minimum. This is to account 
for the well established fact that the easy magnetization direction lies in the plane of 
the FePc molecule.\cite{Barraclough,Barto10} 

A further experimental fact to take into consideration is that the ground state is 
a spin triplet ($S=1$) and that is is orbitally degenerate ($^3E_g$).\cite{Filoti,Barto10}
Within the sector defined by the condition (\ref{condition}) there are only two domains
where $^3E_g$ is the ground state --- a quadrangle $T_3$ and a triangle $T_5$. We carried 
out an extensive numerical study of the magnetic susceptibility (with due allowance for 
the spin-orbit coupling) and found that $\chi_{||}(T) > \chi_{\perp}(T)$ everywhere within 
$T_3$, but $\chi_{||}(T) < \chi_{\perp}(T)$ inside $T_5$. (Here the subscript "$||$" refers
to the direction parallel to the 4-fold symmetry axis.) One has to conclude, therefore, 
that the standpoint of FePc in Figure \ref{analytic} lies inside the triangle $T_5$.
The corresponding ground-state configuration is $a_{1g}^2 e_g^{\uparrow\downarrow\uparrow}
b_{2g}^{\uparrow}$, cf. Eq. (T5). It is distinct from the configuration $T_3$, or
$b_{2g}^2 e_g^{\uparrow\downarrow\uparrow} a_{1g}^{\uparrow}$, postulated by Dale et
al.\cite{Dale} and adopted by Filoti et al.\cite{Filoti} On a simple model the latter
authors have demonstrated that $T_3$ has necessarily an easy-axis anisotropy, which agrees
with our analysis. The experiment,\cite{Barraclough,Barto10} however, insists on an
easy-plane anisotropy and so $T_3$ has to be definitively abandoned. After all, Dale's 
choice of $T_3$ was a mere conjecture, without a sufficient experimental foundation. 
It should also be noted that Miedema et al.\cite{Miedema,Stepanow} proceeded from the 
correct ground-state configuration $T_5$, even though they did not explain their choice.

\begin{figure}
\centerline{\includegraphics[width=0.45\textwidth]{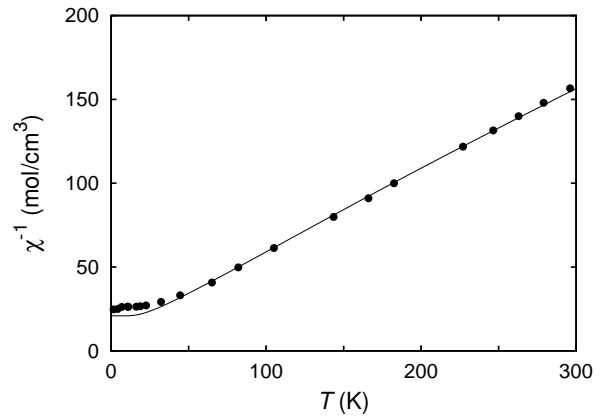}}
\caption{\label{chi} Temperature dependence of reciprocal susceptibility. Closed circles are 
experimental data,\cite{Barraclough} solid line is 1.22 times the calculated $\chi^{-1}$. }
\end{figure}

The difference between the two $^3E_g$ configurations is easy to understand.
In both cases there is one $e_g$ hole; the two real $e_g$ orbitals ($d_{xz}$
and $d_{yz}$) can be combined to give states with $\ell_z=\pm 1$. An extra singly
occupied orbital in $T_3$ is
$a_{1g}$ ($d_{z^2}$), with $\ell_z=0$. Therefore, the z-component of the total
orbital moment is $L_z=\pm 1$ and the spin-orbit coupling leads to an easy-axis
anisotropy in $T_3$. In $T_5$ it is the $d_{xy}$ orbital ($b_{2g}$) that is
singly occupied and the situation is quite different. Now three orbitals,
$d_{xy}$, $d_{xz}$, and $d_{yz}$, are accessible to the holes. If the $e_g$ and
$b_{2g}$ levels were perfectly degenerate, there would be no anisotropy at all.
The fact that the degeneracy is lifted results in a weak easy-plane anisotropy,
such as the one observed.

We undertook an attempt to refine the position of the system inside the triangle $T_5$
on the basis of the available susceptibility data.\cite{Barraclough} We find that
the most likely standpoint is near the left corner of the triangle, at $B_{20}/C=0.84$,
$B_{40}/C=0.0074$. Powder susceptibility was calculated as $\frac{1}{3}\chi_{||} +
\frac{2}{3}\chi_{\perp}$, with $B=917\,{\rm cm}^{-1}$ and $C=4040\,{\rm cm}^{-1}$
(as in Table 7.3 of Ref. \onlinecite{AB}). The spin-orbit coupling constant $\zeta$
was set to $400\,{\rm cm}^{-1}$. The so computed susceptibility proved higher than
the experimental one and had to be reduced by a factor of 0.8, to make both curves 
match. (Accordingly, in Fig. \ref{chi} the experimental reciprocal 
susceptibility\cite{Barraclough} is compared with the calculated $\chi^{-1}$ times 1.22.)
The reduction factor 0.8 can be attributed to covalency, neglected in our model.

Apart from the rescaling, the calculated $\chi^{-1}(T)$ does agree with the experiment.
In our calculation the sextet $^3E_g$ is split by the spin-orbit interaction. The ground
state is a singlet and so is the first excited state, situated $20\,{\rm cm}^{-1}$ above
the ground state. The second excited state, at $52\,{\rm cm}^{-1}$, is a doublet, followed
by two singlets, at $165\,{\rm cm}^{-1}$ and $225\,{\rm cm}^{-1}$. It will be recalled that
the model spectrum of Refs. \onlinecite{Dale} and \onlinecite{Barraclough} consisted of
a ground singlet and an excited doublet at $64\,{\rm cm}^{-1}$. The most essential 
distinction of our spectrum is the presence of an excited singlet at $20\,{\rm cm}^{-1}$.
A clue to this point might be provided by a measurement of the specific heat. The isolated
molecule has no magnetic moment but the application of an external magnetic field $H_x$
in easy-plane direction gives rise to a spin moment $m_S^x=-2\mu_{\rm B} \langle \hat S_x \rangle$
that saturates at about $m_S^x \approx 2 \mu_{\rm B}$ for fields exceeding 40 T in agreement
with $S=1$. We find a ratio of orbital and spin moments 
$m_L^x/m_S^x=\langle \hat L_x \rangle / \left( 2 \langle \hat S_x \rangle \right) \approx 0.65$
for our refined parameter set in reasonable agreement with the ratio of 0.83 that was 
measured by XMCD.\cite{Barto10,remark} Therefore, we confirm the existence of an extraordinarily
large, highly unquenched orbital moment in FePc.

\section{Conclusion}
Published experimental data suggest that FePc has an orbitally degenerate ground state
with $S=1$, the easy magnetization direction lying in the plane of the molecule.
There is a single domain in the CF parameter space where these conditions are met ---
the triangle $T_5$ in Figure \ref{analytic}. The corresponding ground-state
configuration is $a_{1g}^2 e_g^3 b_{2g}^1$. The standpoint of FePc is situated in
the left corner of the triangle, about $B_{20}/C=0.84$, $B_{40}/C=0.0074$, whereas $B_{44}$ 
is given by Eq. (\ref{superpos}). This point lies in a strong-CF region, where the notion 
of single-determinant states has a certain validity.

\begin{acknowledgments}
The authors are thankful to Dr. Guillaume Radtke for helpful discussions. A significant 
part of this work was carried out during a three-month stay of M.D.K. at the University
of Aix-Marseille and he wishes to express his gratitude to the staff at the Faculty of
Sciences for hospitality and to CNRS for financial support.
\end{acknowledgments}


\begin{thebibliography}{99}
\bibitem{remark1}
N.B. In some of the more recent literature \cite{MaromKronik,Wang,Sumimoto} 
(but not in Ref.\ \onlinecite{Nakamura}) 
$a_{1g}^2 b_{2g}^2 e_g^{\uparrow\uparrow} $ is erroneously referred to as
$^3A_{1g}$.
\bibitem{remark}
Our absolute value of $m_L^x=1.3 \mu_{\rm B}$ for a field of 40 T exceeds the measured XMCD value of 0.53 
$\mu_{\rm B}$, but for a detailed analysis of the XMCD results also the intra-atomic magnetic dipole operator
has to be taken into account, which is beyond the scope of the present article.
\end{thebibliography}
\end{document}